\documentclass[aip,graphicx, twocolumn]{revtex4-1}
\usepackage{graphicx}
\usepackage{natbib}
\DeclareUnicodeCharacter{2009}{ }

\draft 

\begin{document}
\bibliographystyle{unsrt}

\preprint{}
 
\title{Cloud and Microjet Mix:\\ A Possible Source of Yield limitation of the National Ignition Facility Targets} 



\author{Gene H. McCall}
\email{ghm@lanl.gov}
\affiliation{Los Alamos National Laboratory, Los Alamos, NM 87545}


\date{\today}

\begin{abstract}
The National Ignition Facility(NIF) megajoule laser at the Lawrence Livermore National Laboratory(LLNL)\cite{moses} has produced a significant amount of useful physics results related to inertial confinement fusion since it began operating in 2010. However, achievement of its primary goal of generating ignition in a thermonuclear plasma has proven elusive, and measured yields were far below those expected from an ignited, and burning, plasma. The reason for the limited yields has not been explained in detail. This work proposes, and shows support for, the possibility that the low yields may be the result of ejecta from the interior of the target containment, and driving, shell.
\end{abstract}


\pacs{}

\maketitle 

\section{Introduction}
The \emph{National Ignition Facility(NIF)} is a large neodymium-glass laser covering an area the size of a football stadium at the Lawrence Livermore National Laboratory(LLNL)\cite{moses}. It focuses 192 beams onto a millimeter-scale target at power levels as high as 500 TW and energies up to 1.9 MJ. It was built for the purpose of providing scientific information to support the Department of Energy(DOE) nuclear weapon Science-based Stockpile Stewardship program. Its primary goal is to demonstrate the ignition of a thermonuclear fuel\cite{paisner}. Results have been disappointing with energy yields far below those expected for an ignited, burning thermonuclear plasma. A recent shot, however, has produced a yield of 1.3 MJ\cite{mjshot}. Attempts to reproduce the result, though, produced significantly less energy. Two shots in December, 2021 and January, 2022, gave 700 kJ and 430 kJ, respectively. A second November shot is still being analyzed\cite{lesse}. possible reasons for the discrepancy will be discussed below.

This work is not intended to provide a critique of target design for the NIF, nor is it intended to produce new target designs. Rather, it will show that an important physical effect(ejecta) was ignored during the preparations for NIF operation, and will suggest possible ways to reduce its effect.

\section{Ignition}

ICF ignition has been defined by Lindl\cite{lindl:ign}. The intuitive definition as the onset of a  self-sustaining burn applies equally well to a thermonuclear process and to a college bonfire. It should emphasized that although the 1.3MJ shot represents a significant achievement, ignition did not occur. The LLNL website, which has compelling reasons to extoll the result for programatic and public relations reasons, does not claim that ignition was achieved. It states that the result places the program at \emph{the threshold of ignition}\cite{llweb} 

During ICF ignition, a central fuel region, or \emph{hot-spot}, will ignite, and the burn will propagate, mainly by electron conduction, to the outer fuel region. \cite{lindl:ign} Conduction and radiation losses cool the hot spot and must be compensated by alpha particle deposition. The hot-spot generated in the targets defined by the NIC was calculated to have a mass of, approximately 20 micrograms.

A 1997 review of the ICF program by a committee of the National Academy of Sciences\cite{nas} defined ignition as signified by a thermonuclear energy output equal to the input laser energy. This condition is, though, not an appropriate definition of ICF ignition, but, rather a definition of ICF \emph{breakeven}, or a target gain of unity. Christopherson, Betti, and Lindl have given a more accurate definition of ignition as the initiation of a thermal instability in a DT plasma, and they describe the inaccuracy of the Academy definition\cite{chris}. The Academy definition was defined as \emph{breakeven} early in the ICF program, and was established as the definition of programmatic success. It became, thus, a popular, and publicized, quantity with little scientific support. Fortunately, in this work, detailed descriptions of targets may be cited, but the results will not depend strongly on them. It appears that the difficulty in defining ignition was not a difficulty in defining the appropriate physical processes, but, rather,defining the observables that signified ignition.

The plasma conditions necessary, and sufficient, for the onset of ignition have been described in many places. A good description is given by Atzeni and Meyer-ter-Vehn\cite{atzeni}. They note that ignition occurs when the energy loss from the target is overcome by the deposition of alpha particles from thermonuclear reactions. The primary loss mechanism they address is bremsstrahlung emission. They calculate an \emph{ideal ignition temperature} of 4.3 keV, assuming that all the alpha particles are stopped in the fuel. This value is consistent with those calculated by the target design codes\cite{lindl:ign}.
\section{NIF Ignition Targets}

There are two categories of NIF targets that were calculated to ignite which are appropriate for consideration in this work. One is the point design target that was the result of the National Ignition Campaign\cite{haan}(NIC), and the other is the HYBRID-E target that produced the highest yield yet achieved on the NIF\cite{krit}.

\begin{figure}[h]
\includegraphics[width=3.2  in.]{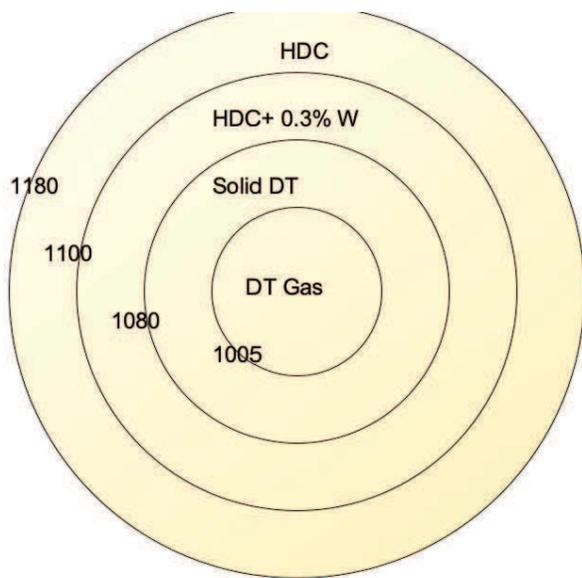}
\caption{Diagram of the HYBRID-E target. Dimensions are the inner radii of concentric spheres in micrometers. Not to scale. DT is a deuterium-tritium mixture, and HDC is high density carbon.
(nanocrystalline diamond)}
\label{fig:he}
\end{figure}

A diagram of the HYBRID-E target is shown in fig. \ref{fig:he}.

\begin{figure}[h]
\includegraphics[width=3.2 in.]{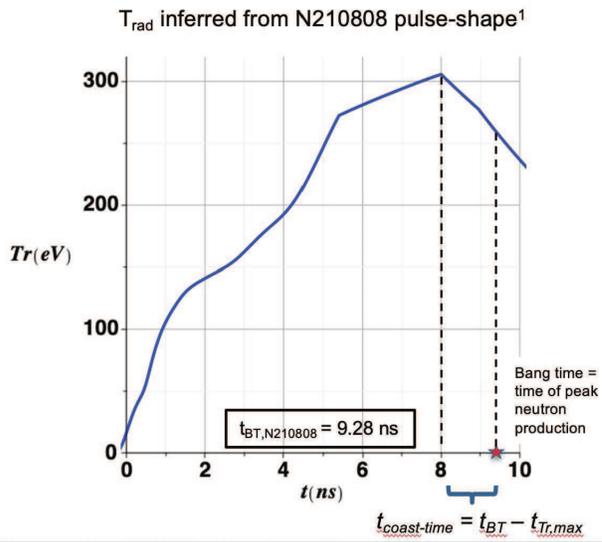}
\caption{Radiation profile for the 1.3 MJ shot\cite{hurricane}}
\label{fig:big}
\end{figure}

The radiation profile driving the target in the 1.3 megajoule output\cite{hurricane} shot is shown in fig. \ref{fig:big}. The radiation profile was inferred from the measured laser intensity, a method known to be accurate.
The target of fig.\ref{fig:he} was modeled with a one-dimensional calculation using the LASNEX\cite{zim} code. The calculated yield was 13.9 MJ.

\begin{figure}
\includegraphics[width=3.2 in.]{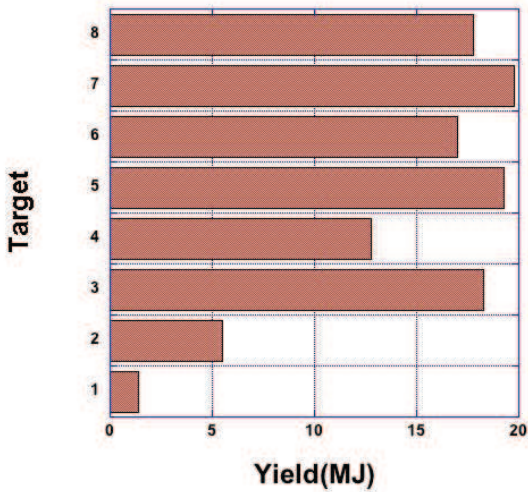}
\caption{predicted yields of various point design targets\cite{haan}.}
\label{fig:ptdes}
\end{figure}

The point design target\cite{haan} was, actually, a set of eight targets having different ablator materials and producing different predicted yields. A bar graph showing the eight predicted yields corresponding to the columns of table I in the point design reference\cite{haan} is shown in fig. \ref{fig:ptdes}. Seven of the eight designs were predicted to have a yield greater than 5 MJ, and the lowest yield target was predicted to have a yield of 1.4 MJ.

\section{One dimensional models}

Describing the dynamics of an inertial fusion implosion is a very complex process, and computer codes in two- and, even, three-dimensions are now being used to model the system. One-dimensional calculations\cite{zim} are still, however, useful in describing general physical characteristics of the implosion. It should be remembered that the original description of inertial fusion physics was based on one-dimensional calculations\cite{nuck}. Too, there is said to be a significant program component dedicated to the design of targets that can be described by one-dimensional models\cite{kline}.Thus, it appears that one dimensional calculations continue to be useful for describing ICF physics, and that they will continue to find applications well into the future.

\section{Mix}

\begin{figure}[h]
\includegraphics[width=3.2 in.]{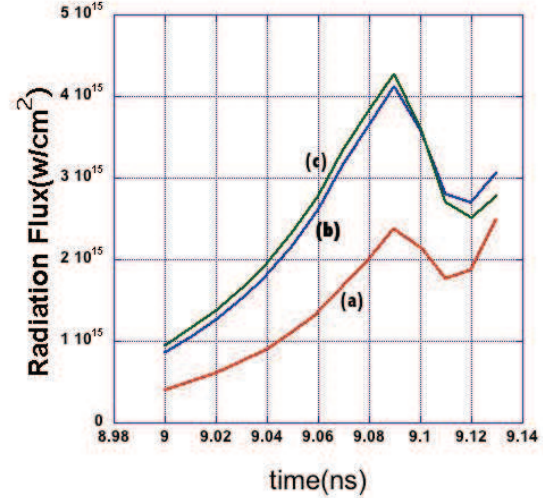}
\caption{Radiation loss,W/cm$^2$} as a function of time during the implosion. (a) is for no mix, (b) is for 3 micrograms of mix, and (c) is for 4 micrograms of mix
\label{fig:mixrad}
\end{figure}

It is well-known that mix is the injection of shell material into the thermonuclear fuel. The foreign material can result in higher radiation losses and increased fuel heat capacity. Those losses must be overcome by the thermonuclear burn if ignition is to occur.  A LASNEX calculation of the radiation loss from a HYBRID-E target as a function of time during the implosion is shown for several amounts of high density carbon plus 0.03\% by atom of tungsten, where the mix, assumed to be deposited uniformly in the fuel, is shown in Fig, \ref{fig:mixrad}. The increasing loss with increasing mix mass can be seen to be substantial. There is a valid question as to the effect of the mix distribution in the fuel. It should be noted that the velocity of the mix material is higher than the inner surface velocity of the shell. as described below and the mix deposition will occur before there is significant motion of either the shell or the fuel. The calculations were, also, done with the mix deposition apportioned according to the initial relative rho-R of the solid and gaseous fuel regions. That method is consistent with the deposition being controlled by aerodynamic stopping of the mix particles. It was found that both methods gave the same result. It is possible, of course that the deposition of the mix material is more complex than either of the approximations used, but that is beyond the scope of this work. That issue can be addressed if experiments show the effect to be worth further study.

It appears that the first example of microjet mix, although not identified as such, was described by Pak and collaborators\cite{pak}, who compared 3-D calculations to experiments which varied the diameter of the fill tube used to load the fuel gas into the target. It was shown that radiation loss from the shell snd tube material injected into the target hot spot reduced the yield. There is also a mention of mix produced by localized capsule imperfections.The reduction produced by the shell mix in the targets described here was somewhat larger because of the presence of tungsten in the HDC shell.

Most of the reports on the performance of the point design targets\cite{haan} express mix in terms of the mass of mix in the \emph{hot spot}. For this work, however, it is appropriate to express the mix in terms of an initial mix that is uniform in the solid DT. Typical hot spot mass is, approximately 20\% of the total fuel mass, and for approximate comparison purposes, the quoted mix masses are multiplied by a factor of five to compare to results for the HYBRID-E target with uniform fuel mix .Because of the ways in which the point design mix mass is used in this work, the error is not significant. It is clear, however, that the HYBRID-E target is more tolerant of mix than are the point design targets.

\begin{figure}[h]
\includegraphics[width= 3.2 in.]{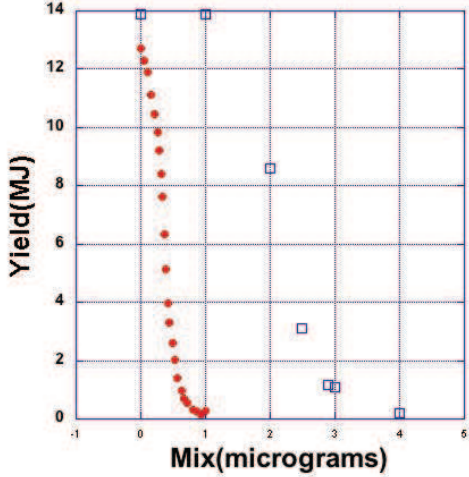}
\caption{Comparison of yields of the NIC point design target and the HYBRID-E target as a function of mix mass. Closed circles are for the point design, and open squares are for the HYBRID-E.}
\label{fig:mixcomp}
\end{figure}

The yield of the HYBRID-E target was calculated using LASNEX for several values of mix mass deposited uniformly in the solid DT. A comparison of those results to calculations of the point design target  is shown in fig. \ref{fig:mixcomp}.

An exponential fit to the HYBRID-E result for mix masses greater than 2 $\mu g$ agrees with the calculated yield to within 12 percent.
The fitting function is: $$ Y= a \times exp(-b M) $$ where Y is the yield in MJ, a=308.99, b=1.1528, and M is the mix mass in $\mu$g.

A yield of 1.3 MJ corresponds to, 2.95 micrograms of mix. The700 kJ shot corresponds to 3.29 $\mu$g of mix, an increase of only12 percent, and the 430 kJ shot corresponds to a mix mass of 3.55 $\mu g$, another increase of only 8 percent\cite{lesse}. Thus, rather large changes in yield can be caused by only small changes in mix mass

Experiments done at the NIF have measured mix in ignition scale targets by doping the target ablator shell with an x-ray emitting impurity. In one case the inner layer of the ablator was doped with copper, and the outer layer was doped with germanium\cite{regan}. Hot spot mix masses of 34 ng to  $4  \mu$g were measured. Those masses are consistent with those discussed above, which were required to produce the observed yields. It was very interesting in that the spectral measurements showed that germanium from the outer layer appeared in the central fuel region, but there was no copper.

It appears that the amount of mix required to reproduce the NIF results has been identified. It remains to identify the most likely cause of the mix, and, if possible, to define ways of eliminating it.

\section{Possible causes of NIF mix}

There are several possible sources for mix. it will be shown that the most popular processes may not be the most likely ones.
Thr review of the National Ignition Campaign by Lindl, et al\cite{lindl-nic}, provides a good discussion of calculations and experiments related to explaining the limited yields from NIF targets. They note that mix material can be generated by the fill tube used to load the fuel into the target. They present calcultions where it is calculated that the mix mass produced is 30 ng, and is premixed uniformly in the hot spot. They note that:
\begin{quotation}
However, the large spread in yields and in the fuel ion temperatures for otherwise similar implosions is not yetpredicted by the 3D calculations. This spread in yield and ion temperature can be largely explained by variability in the hot spot mix observed in these implosions.
\end{quotation}
This statement is consistent with the mix source described below.

\subsection{Fluid instability}

If one is calculating target performance with a fluid dynamics code, as is always done for ICF targets, the processes that could cause the penetration of one fluid element into another, that immediately come to mind, are the Rayleigh-Taylor(RT) and Richtmyer-Meshkov(RM) instabilities\cite{dimonte}. However, during the NIC, large amounts of modeling effort and target fabrication engineering were  spent to eliminate this possibility. Apparently, the work was successful. It has been observed that to force the experiments to agree with models, unphysical multipliers of 3-5 applied to the actual target surface roughness are required.\cite{smal}.

Of course, there is the possibility that misalignment or poor power balance among the 192 beams is producing asymmetries\cite{scott}, but,given the attention and care devoted to beam balance before each shot, it is unlikely that all of the failed shots have such a problem.

Therefore, it appears that the R-T and R-M instabilities are not important effects limiting the yield of all NIF targets. In the discussion below, a more likely cause will be discussed.

\subsection{Microjets and Clouds}

The terms, microjets and clouds are used in this work to distinguish between two manifestations of ejecta emitted when a shock emerges at a free surface. The phenomenon has, probably, been studied since the 1950s, but one of the first journal publications was a description of  powder-gun-driven experiments done at the Sandia National Laboratory\cite{asay-1}. Earlier work is described in ref. \cite{phermex}. A review is given by Buttler\cite{buttler}.Many experiments have been driven by high explosives, but laser-driven shocks have been used\cite{saunders-1}.Many of the experiments on ejecta have used grooves machined on the surface of a material to compare with model calculations as described by Asay and Bertholf\cite{asay-2}. The general characteristic of ejecta formation have been summarized by Zellner, et al\cite{mbz}. They observe that when the shock pressure exceeds the yield strength of the material, imperfections in the shock output surface form microjets. The approximate mass of the jets is the mass required to fill the imperfections at the original density of the material. The so-called \emph{missing mass}

If the surface were completely smooth, though, there would still be ejected particles. Most metals and many non-metals are composed of grains that are loosely bound together. When the materials melt, the grains can separate, and grain-size particles will be ejected. This effect was seen clearly in the experiments of McMillan and Whipkey\cite{mcm} where a shock emerged from a Tin surface said to be smooth to 0.2 $ \mu$m and emitted particles as large as 50 $\mu$m were observed along with many particles of diameters of order 10 $\mu$m. Shock emergence from a HDC shell, as used in NIF experiments will produce clouds of diamond nanocrystals having sizes characteristic of the grain size\cite{nano} with diameter ranging from a few to 100 nm.
The point to be taken here is that an ICF ablator shell containing DT fuel will always generate fuel mix if a shock producing a pressure greater than the melting pressure of the shell propagates from the shell into the fuel. There will be some reduction in the mass of the shell material injected because the fuel region is not a vacuum, but even solid DT has a density of only 0.25 g/$cm^3$, and the retarding force will be small.
The high explosive results, such as particle size distributions\cite{size}, are said to be described by 3D percolation theory.\cite{perc}.

Although the usual procedure in a work such as this would be to present code calculations showing the generation of microjets from typical material surface imperfections, that will not be done here. Actual, and accurate, dimensions of imperfections occurring in actual ICF targets are not known for targets of interest here, only that some exist. And, there are, literally, hundreds of such calculations documented in the literature. A simple search for ejecta calculations using \emph{Google Scholar} will produce links to them. Further generalized calculations of that type would simply be redundant and space consuming. Where appropriate, specific references will be given. Experiments, not more calculations, are needed.

An extensive literature search and communication with a recognized expert in the field of ejecta physics, failed to discover a single case where an experiment which produced shock conditions as described above, failed to generate either microjets, particle clouds, or both. There have been some attempts to vary experimental geometries to suppress ejecta, with limited results\cite{georges}.

Thus, it appears correct to say that absence of ejecta in the fuel of a Hybrid-E target would be a physical rarity, and HDC mix material in the fuel is certain to occur.The quantity of mix for the typical target materials and configurations  used at the NIF should be determined by experiment.

LASNEX calculations using the drive shown in fig.\ref{fig:big} predict that the first shock energing from the HDC shell has a pressure below the melt pressure of HDC, but still above the yield strength. Microjets will be produced if there are surface imperfections in the interior surface of the shell.The second shock presure, however, is well above the melt pressure, and a cloud of nanocrystals will be projected into the fuel.

\section{High Density Carbon}

High density carbon(HDC) is, currently, the ablator material of choice for NIF targets\cite{krit}.The material\cite{hdcfab} is composed of diamond nanocrystals, bound together at their edges,and it is, frequently, identified as \emph{Poly-nanocrystalline diamond (PND)}The material not only improves the hydrodynamic coupling between the radiation field and the target, but it has mechanical properties that can be used to reduce fuel energy losses and increase yield. These properties play a large part in the discussion to follow. 

\subsection{Shock Response}

\begin{figure}[h]
\includegraphics[width=3.2 in.]{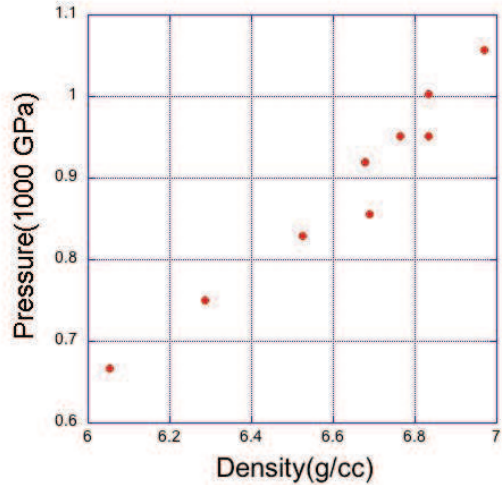} 
\caption{Melt hugoniot of high density carbon. Melting begins at, approximately, 6 MBar, and is complete at, approximately, 7 MBar.}
\label{fig:knud}
\end{figure}

The static\cite{philip} and dynamic\cite{katagiri} properties of HDC have been investigated in detail.The shock melting curve measured by Knudson\cite{knudson} is shown in fig. \ref{fig:knud}.
The high melting pressure of HDC could, possibly, reduce the total number of particles injected into the fuel region, but, unfortunately, the initial shock generated by the drive of fig, \ref{fig:big} is greater than the yield pressure of the HDC ablator.

The mean size of of the nanocrystals in a HDC layer has been measured to be 55 nm $\pm$20 nm\cite{nano}. The inferred mix of 2.8 $\mu$g corresponds to nearly $10^{10}$ particles injected into the fuel, but corresponds to an inner layer of the HDC shell only 54 nm thick. Thus, the material needed to produce the effects calculated require an ejected layer only as thick as the diameter of one nanocrystal\cite{nano}.The amount of ejected mass is consistent with the roughness associated with the shell fabrication\cite{biener}.

Independent of whether one considers the numbers involved with the mix mass to be large, or small, the fact remains that a target driven by an initial shock is unlikely to ignite. The question to be answered is whether a radiation driving field can be devised such that an initial strong shock does not occur. 
This possibility will be discussed in the next section.

\section{Shockless Drive}

The possibility of shockless, or compressive, drive has been studied from the point of view of maintaining a low adiabat in ICF targets. Both indirect drive\cite{xue} and direct drive\cite{swift} have been addressed.

Before attempting a full target shot with a new driving profile, however, experiments using a sample of the HDC ablator material should be tested in the manner described by Saunders\cite{saunders-2} to determine the magnitude of the microjet and cloud problem.

In this work an attempt to produce compressive drive was attempted using  a radiation drive field that rose linearly in time to a maximum temperature of 325 eV. After reaching the maximum temperature, the yield did not depend on whether the temperature was kept constant, or reduced to zero.Given that the measured compressive sound speed of  HDC is 17,980 m/s\cite{philip} a compression wave should emerge at  time of 5.56 ns after the start of the driving pulse. Linear rise times of 10 ns, or greater appear to satisfy this condition.

\begin{figure}[h]
\includegraphics[width=3.2 in.]{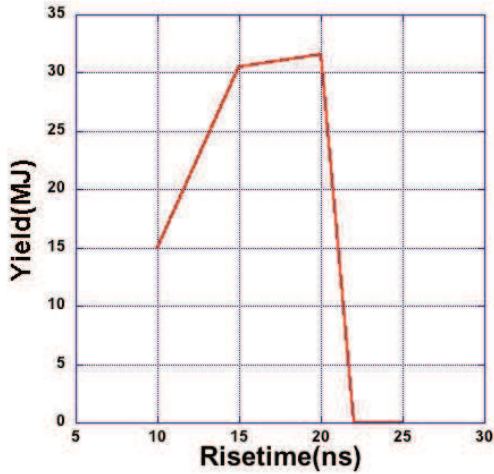}
\caption{Yield of a HYBRID-E target driven by  linearly rising radiation pule as a function of the radiation pulse risetime}
\label{fig:lin}
\end{figure}

One-dimensional LASNEX calculations of the yield produced by a HYBRID-E target driven by a linear radiation pulse for several values of risetime are shown in fig. \ref{fig:lin}. The linear rise is seen to produce ignition and high yield by eliminating mix and by establishing a low adiabat. The target yield is predicted to be as high as 30 MJ.
It remains to be seen whether the NIF can produce such a radiation driving pulse, but, if not, similar options should be explored.

\section{Conclusions}

The HDC shells currently used in the fabrication of NIF targets offer significant possibilities for producing thermonuclear ignition and high yield. It is unlikely that the current driving pulses will cause a target to ignite, but changes in the driving pulse shape to produce compressive drive are likely to succeed. 
Although the calculations presented here were done in only one dimension, the possibilities presented warrant further theoretical and experimental studies.

\section{Acknowledgements}

The author thanks James Asay, W. T. Buttler, and Mark Schmitt for many significant, and helpful, comments and suggestions. This work was supported by Triad National
Security, LLC, for the National Nuclear Security Administration
of the U.S. Department of Energy under Contract No.
89233218CNA000001.




%
%

%


\end{document}